\documentclass[times,doublespace]{anzsauth}

\usepackage{url}
\usepackage[UKenglish]{isodate}
\usepackage{multirow}
\usepackage{bm}

\newtheorem{theorem}{Theorem}

\def\volumeyear{ }

\begin{document}

\cleanlookdateon
\runningheads{Semi-supervised Gaussian mixture modelling in R}{Z~Lyu, D~Ahfock, R~Thompson, G~J~McLachlan}
\title{Semi-supervised Gaussian mixture modelling with a missing-data mechanism in R}
\author{Ziyang Lyu\addressnum{1}\corrauth, Daniel Ahfock\addressnum{2}, Ryan Thompson\addressnum{1,3} and Geoffrey J. McLachlan\addressnum{2}}

\affiliation{University of New South Wales, University of Queensland and CSIRO's Data61}
\address{
\addressnum{1} UNSW Data Science Hub and School of Mathematics and Statistics, University of New South Wales, NSW 2052 Australia \hspace*{1ex} Email: \texttt{ziyang.lyu@unsw.edu.au} \\
\addressnum{2} School of Mathematics and Physics, University of Queensland, QLD 4072 Australia \\
\addressnum{3} Data61, Commonwealth Scientific and Industrial Research Organisation, NSW 2015 Australia
}

\begin{abstract}
Semi-supervised learning is being extensively applied to estimate classifiers from training data in which not all the labels of the feature vectors are available. We present \textsf{gmmsslm}, an \textsf{R} package for estimating the Bayes' classifier from such partially classified data in the case where the feature vector has a multivariate Gaussian (normal) distribution in each of the predefined classes. Our package implements a recently proposed Gaussian mixture modelling framework that incorporates a missingness mechanism for the missing labels in which the probability of a missing label is represented via a logistic model with covariates that depend on the entropy of the feature vector. Under this framework, it has been shown that the accuracy of the Bayes' classifier formed from the Gaussian mixture model fitted to the partially classified training data can even have lower error rate than if it were estimated from the sample completely classified. This result was established in the particular case of two Gaussian classes with a common covariance matrix. Here, we focus on the effective implementation of an algorithm for multiple Gaussian classes with arbitrary covariance matrices. A strategy for initialising the algorithm is discussed and illustrated. The new package is demonstrated on some real data.
\end{abstract}


\maketitle

\section{Introduction}
\label{sec:intro}

Classifiers such as neural networks often achieve their strong performance through a completely supervised learning approach, which requires a fully classified (labelled) dataset. However, missing labels often occur in practice due to difficulty determining the true label for an observation (the feature vector). For example, in medicine and defence, images can often only be correctly classified by a limited number of experts in the field. Hence, a training sample might not be completely classified, with images difficult to classify left without their class labels. Moreover, in medicine, there might be scans that can be diagnosed confidently only after an invasive procedure, possibly regarded as unethical to perform at the time.

Semi-supervised learning \citep{chapelle2006semi} addresses the issue of missing labels. Classic approaches that belong to the semi-supervised learning paradigm include generative models \citep{pan2006semi,kim2007texture,fujino2008semisupervised}, graph-based models \citep{blum1998combining,szummer2001partially,zhou2003learning}, and semi-supervised support vector machines \citep{vapnik1998support,joachims1999transductive,lanckriet2004learning}. Gaussian mixture models are a fundamental class of statistical models particularly relevant to semi-supervised learning. Given a partially classified sample, the traditional optimisation objective is a joint likelihood over the labelled and unlabelled data. This problem of maximum likelihood (ML) with missing data is amenable to the expectation-maximisation algorithm of \cite{dempster1977maximum}; see, for example, \cite{mclachlan2000finite}, \cite{mclachlan2008algorithm} and the recent review of \cite{mclachlan2019finite} on finite mixture models. Although Gaussian mixture models in the semi-supervised setting are now well studied \citep{pan2006semi,kim2007texture,come2009learning,huang2010semi,szczurek2010introducing}, there is often a critical assumption that the missing-label process can be ignored for likelihood-based inference \citep{mclachlan1975iterative,mclachlan1977estimating,ganesalingam1978efficiency,chawla2005learning}.

Recently, \cite{ahfock2020apparent} introduced a novel approach that treats labels of unclassified observations as missing data, leveraging a framework for handling missingness, as in the groundbreaking work by \citet{rubin1976inference} on incomplete data analysis. \citet{ahfock2020apparent} conducted an asymptotic analysis to demonstrate that a partially classified sample can provide more valuable information than a fully labelled sample, specifically in the two-class Gaussian homoscedastic model; see also the review by \citet{ahfock2023semi}. By building upon their framework, we propose to model the probability of a missing label as a logistic regression model where the mean depends on a single covariate equal to an entropy-based measure. This approach enables the implementation of an algorithm for estimating the Bayes' classifier through the full likelihood, catering to multiple Gaussian classes with arbitrary covariance matrices. Given the complexities that emerged in the derived expressions, transitioning from the specific two-class homoscedastic Gaussian model to this expanded framework poses a significant technical challenge. Our development and presentation of an \textsf{R} package designed for semi-supervised learning from a statistical viewpoint marks an important advancement. Its potential spans various statistical analyses and application scenarios.

In the context of our more general framework, we introduce the \textsf{R} package \textsf{gmmsslm} (\underline{G}aussian \underline{m}ixture \underline{m}odel-based \underline{s}emi-\underline{s}upervised \underline{l}earning with a \underline{m}issing-data mechanism), which is available open-source on the Comprehensive \textsf{R} Archive Network at \url{https://cran.r-project.org/package=gmmsslm}. This package implements three distinct Gaussian mixture modelling approaches: (1) partially classified samples, taking into account the missing-data mechanism; (2) partially classified samples, disregarding the missing-data mechanism; and (3) fully classified samples. Although various packages exist for estimating mixture models, such as \textsf{bgmm} \citep{biecek2012r}, \textsf{EMMIX} \citep{mclachlan1999emmix}, \textsf{flexmix} \citep{grun2007flexmix}, \textsf{mclust} \citep{fraley2007model}, \textsf{mixtools} \citep{benaglia2009mixtools} and \textsf{Rmixmod} \citep{lebret2015rmixmod}, none accommodate a missing-data mechanism. In \textsf{gmmsslm}, the missingness mechanism is specified through a multinomial logistic regression concerning the entropy of feature vectors. The package applies to an arbitrary number of classes possessing multivariate Gaussian distributions with potentially dissimilar covariance matrices.

This paper is structured as follows. Section~\ref{sec:gaussian} provides a concise review of the statistical model used in \textsf{gmmsslm}. Section~\ref{sec:package} describes how users can implement \textsf{gmmsslm} in applications. Section~\ref{sec:example} illustrates a practical application of the package. Finally, Section~\ref{sec:summary} concludes the paper.

\section{Gaussian mixture model with a missing-data mechanism}
\label{sec:gaussian}

\subsection{Gaussian mixture model}

We let $\bm{Y}$ be a $p$-dimensional vector of features on an entity to be assigned to one of $g$ predefined classes $C_1,\ldots,C_g$ occurring in proportions $\pi_1,\ldots,\pi_g$, where $\sum_{i=1}^g\pi_i=1$. The random variable $\bm{Y}$ corresponding to the realisation $\bm{y}$ is assumed to have density $f_i(\bm{y};\bm{\omega}_i)$ with a vector $\bm{\omega}_i$ of unknown parameters in Class $C_i$ $(i=1,\ldots,g)$. The vector of all unknown parameters is given by $\bm{\theta}=(\pi_1,\ldots,\pi_{g-1},\bm{\omega}_1^\top,\ldots,\bm{\omega}_g^\top)^\top$.

In the sequel, we assume that the class-conditional densities of $\bm{y}$ are multivariate Gaussian with
\begin{equation*}
f_i(\bm{y};\bm{\omega}_i)=\phi(\bm{y};\bm{\mu}_i,\bm{\Sigma}_i)\quad(i=1,\ldots,g),
\end{equation*}
where $\phi(\bm{y};\bm{\mu},\bm{\Sigma})$ denotes the $p$-variate Gaussian density function with mean $\bm{\mu}$ and covariance matrix $\bm{\Sigma}$. The vector $\bm{\theta}$ of all unknown parameters now consists of the elements of the means $\bm{\mu}_i$ and the $p(p+1)/2$ distinct elements of the covariance matrices $\bm{\Sigma}_i$, along with the mixing proportions. In order to estimate $\bm{\theta}$ it is customary in practice to have available a training sample. We let $\bm{x}_{\rm CC}=(\bm{x}_1^\top,\ldots,\bm{x}_n^\top)^\top$ contain $n$ independent realisations of $\bm{X}=(\bm{Y}^\top,Z)^\top$ as the completely classified training data, where $Z$ denotes the class membership of $\bm{Y}$, being equal to $i$ if $\bm{Y}$ belongs to Class $C_i$ $(i=1,\ldots,g)$ and zero otherwise, and where $\bm{x}_j=(\bm{y}_j^\top,z_j)^\top$ $(j=1,\ldots,n)$. For a partially classified training sample $\bm{x}_{\rm PC}$ in semi-supervised learning, we introduce the missing-label indicator $m_j$ which equals 1 if $z_j$ is missing and 0 if it is available $(j=1,\ldots,n)$. Thus, $\bm{x}_{\rm PC}$ consists of those observations $\bm{x}_j$ in $\bm{x}_{\rm CC}$ with $m_j=0$, but only the feature vector $\bm{y}_j$ in $\bm{x}_{\rm CC}$ if $m_j=1$ (that is, the label $z_j$ is missing). The presence of unclassified feature observations in the training data (that is, features with missing labels) necessitates the consideration and fitting of the unconditional density of $\bm{Y}$, which is given by the $g$-component Gaussian mixture density,
\begin{equation*}
f(\bm{y}_j;\bm{\theta})=\sum_{i=1}^g \pi_i\phi(\bm{y}_j;\bm{\mu}_i,\bm{\Sigma}_i).
\end{equation*}

The optimal (Bayes') rule of allocation $R(\bm{y};\bm{\theta})$ assigns an entity with feature vector $\bm{y}$ to Class $C_k$, that is, $R(\bm{y};\bm{\theta})=k$ if $k=\arg\max_i\tau_i(\bm{y};\bm{\theta})$, where $\tau_i(\bm{y};\bm{\theta})=\pi_if_i(\bm{y};\bm{\omega}_i)/\sum_{h=1}^g\pi_hf_h(\bm{y};\bm{\omega}_h)$ is the posterior probability that the entity belongs to Class $C_i$ given $\bm{Y}=\bm{y}$ $(i=1,\ldots,g)$.

We define the log likelihoods
\begin{eqnarray}
\log L_\mathrm{C}(\bm{\theta})&=&\sum_{j=1}^n(1-m_j)\sum_{i=1}^gz_{ij}\log\{\pi_if_i(\bm{y}_j;\bm{\omega}_i)\}, \label{eq:loglik_c} \\
\log L_\mathrm{UC}(\bm{\theta})&=&\sum_{j=1}^nm_j\log\left\{\sum_{i=1}^g\pi_if_i(\bm{y}_j;\bm{\omega}_i)\right\}, \notag \\
\log L_\mathrm{PC}^{(\mathrm{ig})}(\bm{\theta})&=&\log L_{\mathrm{C}}(\bm{\theta})+\log L_{\mathrm{UC}}(\bm{\theta}). \label{eq:loglik_cc}
\end{eqnarray}
In \eqref{eq:loglik_c}, $z_{ij}=1$ if $z_j=i$ and otherwise $z_{ij}=0$. If one ignores the `missingness' of the class labels, $L_\mathrm{C}(\bm{\theta})$ and $L_\mathrm{UC}(\bm{\theta})$ are the likelihood functions formed from the classified and unclassified data, respectively. The likelihood function $L_\mathrm{PC}^{(\mathrm{ig})}(\bm{\theta})$ is formed from the partially classified sample $\bm{x}_\mathrm{PC}$, ignoring the missing-data mechanism for the labels. The likelihood $L_\mathrm{CC}(\bm{\theta})$ for the completely classified sample $\bm{x}_\mathrm{CC}$ is recovered from \eqref{eq:loglik_c} by taking all $m_j=0$.

\subsection{Missing-data mechanism}

In the present context, it is appropriate to dispense with the missing-data mechanism when performing likelihood inference in situations where the missing labels can be viewed as missing completely at random (MCAR) in the framework proposed by \cite{rubin1976inference} for handling missingness in incomplete data analysis. The reader is referred to \cite{mealli2015clarifying} for precise definitions of MCAR and its less restrictive version of missing at random (MAR). The MCAR case here holds if the missingness of labels is independent of both the features and labels, while with the MAR case this missingness is allowed to depend on the features but not the labels. As highlighted by \cite{mclachlan1989mixture}, one can legitimately ignore missingness in certain MAR situations, such as when dealing with truncated features. However, for the MAR scenario being discussed here, one cannot ignore the missingness in carrying out the likelihood analysis. Indeed the use of the missingness provides a way of improving the performance of the classifier to be formed.

\cite{ahfock2020apparent} noted that it is common in practice for unlabelled images (that is, the features with missing labels) to fall in regions of the feature space where there is class overlap. This finding led them to argue that the unlabelled observations can carry additional information that can be used to improve the efficiency of the parameter estimation of $\bm{\theta}$. Additional theoretical motivation is available in Appendix~\ref{app:theory}. They noted that in these situations the difficulty of classifying an observation can be quantified using the Shannon entropy of an entity with feature vector $\bm{y}$, which is defined by 
\begin{equation*}
e(\bm{y}_j;\bm{\theta})=-\sum_{i=1}^g\tau_i(\bm{y}_j;\bm{\theta})\log\tau_i(\bm{y}_j;\bm{\theta}).
\end{equation*}
Let $M_j$ denote the random variable corresponding to the realised value $m_j$ of the missing-label indicator for the observation $\bm{y}_j$. The missingness mechanism of \citet{rubin1976inference} is specified in the present context as
\begin{equation*}
\Pr\{M_j=1\mid\bm{y}_j,\bm{z}_j\}=\Pr\{M_j=1\mid\bm{y}_j\}=q(\bm{y}_j;\bm{\theta},\bm{\xi}),
\end{equation*}
where $\bm{\xi}=(\xi_0,\xi_1)^\top$ is distinct from $\bm{\theta}$. The conditional probability $q(\bm{y}_j;\bm{\theta},\bm{\xi})$ is taken to be a logistic function of the Shannon entropy $e(\bm{y}_j;\bm{\theta})$, yielding
\begin{equation}
\label{eq:qfun}
q(\bm{y}_j;\bm{\theta},\bm{\xi})=\frac{\exp\{\xi_0+\xi_1\log e(\bm{y}_j;\bm{\theta})\}}{1+\exp\{\xi_0+\xi_1\log e(\bm{y}_j;\bm{\theta})\}}.
\end{equation}
In the special case of $g=2$ with equal covariance matrices, the negative log entropy in the conditional probability \eqref{eq:qfun} can be replaced by the square of the discriminant function. Details can be found in Appendix~\ref{app:discriminant}.

\subsection{Full likelihood function based on a missingness mechanism}

We let $\bm{\Psi}=(\bm{\theta}^\top,\bm{\xi}^\top)^\top$ be the vector of all unknown parameters. Henceforth, we let $f$ be a generic symbol for a density or probability function where appropriate. To construct the full likelihood function $L_\mathrm{PC}^{(\mathrm{full})}(\bm{\Psi})$ from the partially classified sample $\bm{x}_\mathrm{PC}$, we need expressions for
\begin{equation*}
f(\bm{y}_j,\bm{z}_j,m_j=0)\quad\text{and}\quad f(\bm{y}_j,m_j=1),
\end{equation*}
corresponding to the classified and unclassified observations $\bm{y}_j$, respectively. For a classified observation $\bm{y}_j$, it follows that
\begin{equation*}
\begin{split}
f(\bm{y}_j,\bm{z}_j,m_j=0)&=f(\bm{z}_j)f(\bm{y}_j\mid\bm{z}_j)\Pr\{M_j=0\mid\bm{y}_j,\bm{z}_j\} \\
&=\prod_{i=1}^g\{\pi_if_i(\bm{y}_j;\bm{\omega}_i)\}^{z_{ij}}\{1-q(\bm{y}_j;\bm{\theta},\bm{\xi})\},
\end{split}
\end{equation*}
while for an unclassified observation $\bm{y}_j$, we have
\begin{equation*}
f(\bm{y}_j,m_j=1)=f(\bm{y}_j)\Pr\{M_j=1\mid\bm{y}_j\}=\sum_{i=1}^g \pi_if_i(\bm{y}_j;\bm{\omega}_i)q(\bm{y}_j;\bm{\theta},\bm{\xi}).
\end{equation*}
The full likelihood can then be expressed as
\begin{equation*}
L_\mathrm{PC}^{(\mathrm{full})}(\bm{\Psi})=\prod_{j=1}^n\{f(\bm{y}_j,\bm{z}_j,m_j=0)\}^{1-m_j}\{f(\bm{y}_j,m_j=1)\}^{m_j}.
\end{equation*}
Thus, the full log likelihood follows as
\begin{equation}
\label{eq:lik_full}
\log L_{\mathrm{PC}}^{(\mathrm{full})}(\bm{\Psi})=\log L_{\mathrm{PC}}^{(\mathrm{ig})}(\bm{\theta})+\log L_{\mathrm{PC}}^{(\mathrm{miss})}(\bm{\Psi}),
\end{equation}
where $\log L_{\mathrm{PC}}^{(\mathrm{ig})}(\bm{\theta})$ is given in \eqref{eq:loglik_cc} and
\begin{equation*}
\log L_{\mathrm{PC}}^{(\mathrm{miss})}(\bm{\Psi})=\sum_{j=1}^n\left[(1-m_j)\log\left\{1-q(\bm{y}_j;\bm{\theta},\bm{\xi})\right\}+m_j\log q(\bm{y}_j;\bm{\theta},\bm{\xi})\right]
\end{equation*}
is the log likelihood formed on the basis of the missing-label indicators $m_j$.

Using the expectation conditional maximisation (ECM) algorithm, we obtain the ML estimations $\hat{\bm{\theta}}_\mathrm{CC}$, $\hat{\bm{\theta}}_\mathrm{PC}^{(\mathrm{ig})}$ and $\hat{\bm{\theta}}_\mathrm{PC}^{(\mathrm{full})}$ for $\bm{\theta}$ based on the log likelihoods $\log L_\mathrm{CC}(\bm{\theta})$, $\log L_\mathrm{PC}^{(\mathrm{ig})}(\bm{\theta})$ and $\log L_\mathrm{PC}^{(\mathrm{full})}(\bm{\Psi})$, respectively (see Appendix~\ref{app:ecm} for details). Notice that $\hat{\bm{\theta}}_\mathrm{PC}^{(\mathrm{full})}$ is a subvector of $\hat{\bm{\Psi}}^{(\mathrm{full})}_\mathrm{PC}$. Similarly, we let $R(\hat{\bm{\theta}}_\mathrm{CC})$, $R(\hat{\bm{\theta}}_\mathrm{PC}^{(\mathrm{ig})})$ and $R(\hat{\bm{\theta}}_\mathrm{PC}^{(\mathrm{full})})$ denote the estimated Bayes' rule obtained by plugging in the estimates $\hat{\bm{\theta}}_\mathrm{CC}$, $\hat{\bm{\theta}}_\mathrm{PC}^{(\mathrm{ig})}$ and $\hat{\bm{\theta}}_\mathrm{PC}^{(\mathrm{full})}$, respectively. The overall conditional error rate of the rule $R(\bm{y};\hat{\bm{\theta}}_\mathrm{CC})$ is then
\begin{equation}
\label{eq:err}
\operatorname{err}(\hat{\bm{\theta}}_\mathrm{CC};\bm{\theta})=1-\sum_{i=1}^g\pi_i\Pr\{ R(\bm{y};\hat{\bm{\theta}}_\mathrm{CC})=i\mid\hat{\bm{\theta}}_\mathrm{CC}, Z=i\}.
\end{equation}
The corresponding conditional error rates $\operatorname{err}(\hat{\bm{\theta}}_\mathrm{PC}^{(\mathrm{ig})};\bm{\theta})$ of the rule $R(\bm{y};\hat{\bm{\theta}}_\mathrm{PC}^{(\mathrm{ig})})$ and $\operatorname{err}(\hat{\bm{\theta}}_\mathrm{PC}^{(\mathrm{full})};\bm{\theta})$ of the rule $R(\bm{y};\hat{\bm{\theta}}_\mathrm{PC}^{(\mathrm{full})})$ are defined likewise. The optimal error rate $\operatorname{err}(\bm{\theta})$ follows by replacing $\hat{\bm{\theta}}_\mathrm{CC}$ with $\bm{\theta}$ in \eqref{eq:err}.

\section{Package description}
\label{sec:package}

In this section, we describe the \textsf{gmmsslm} package. The package has been crafted to implement three distinct Gaussian mixture models, which can be fit using the function \texttt{gmmsslm()}. These models utilise different log likelihood functions, namely \eqref{eq:loglik_c}, \eqref{eq:loglik_cc} and \eqref{eq:lik_full}. Furthermore, the package provides several auxiliary functions including the Bayes' rule classifier, \texttt{bayesclassifier()}, conditional error rate, \texttt{erate()} and Shannon entropy, \texttt{get\_entropy()}. The general syntax for using these functions is illustrated below:
\begin{verbatim}
gmmsslm(dat, zm, paralist, xi = NULL, type)
bayesclassifier(dat, p, g, paralist)
erate(dat, p, g, paralist, clust)
get_entropy(dat, n, p, g, paralist)
\end{verbatim}

Details for some of the main command arguments are as follows:
\begin{itemize}
\item \texttt{n}: Number of observations.
\item \texttt{p}: Dimension of observation vector.
\item \texttt{g}: Number of multivariate normal classes.
\item \texttt{dat}: An $n\times p$ matrix where each row represents an individual observation.
\item \texttt{zm}: An $n$-dimensional vector containing the class labels including the missing label denoted as \texttt{NA}.
\item \texttt{pi}: A $g$-dimensional vector for the initial values of the mixing proportions.
\item \texttt{mu}: A $p\times g$ matrix for the initial values of the location parameters.
\item \texttt{sigma}: A $p\times p$ matrix, or a $p\times p\times g$ array, for the initial values of the covariance matrices. The model is fit with a common covariance matrix if \texttt{sigma} is a $p\times p$ covariance matrix, otherwise the model is fit with unequal covariance matrices.
\item \texttt{paralist}: A list containing \texttt{pi}, \texttt{mu} and \texttt{sigma}. When \texttt{paralist} is provided, individual inputs for \texttt{pi}, \texttt{mu} and \texttt{sigma} are not required.
\item \texttt{xi}: A 2-dimensional vector containing the initial values of the coefficients in the logistic function of the Shannon entropy. The default value of \texttt{xi} is \texttt{NULL}, except when \texttt{type = 'full'} is selected.
\item \texttt{type}: One of three types of Gaussian mixture models as follows: \texttt{'full'} fits the model given by \eqref{eq:lik_full} to a partially classified sample on the basis of the full likelihood by taking into account the missing-data mechanism, \texttt{'ign'} fits the model given by \eqref{eq:loglik_cc} to a partially classified sample based on the likelihood that the missing-data mechanism is ignored and \texttt{'com'} fits the model given by \eqref{eq:loglik_c} to a completed classified sample.
\item \texttt{clust}: An $n$-dimensional vector of the class partition.
\end{itemize}

The main function \texttt{gmmsslm()} provides a comprehensive summary output, detailed as follows:
\begin{itemize}
\item \texttt{Likelihood}: Value of the objective likelihood.
\item \texttt{VarianceStructure}: Describes the structure of the covariance matrix associated with the Gaussian mixture model fitting. Possible structures include `A common covariance matrix' and `Unequal covariance matrices'.
\item \texttt{Convergence}: State of convergence.
\item \texttt{Iteration}: Number of iterations executed.
\item \texttt{TotalObservation}: Total number of observations.
\item \texttt{Dimension}: Dimension of the data.
\item \texttt{ModelType}: Type of model fit. The `full' model type considers a partially classified sample based on the full likelihood, accounting for the missing-data mechanism. The `ign' type fits the model to a partially classified sample, dismissing the missing-data mechanism. The `com' type is dedicated to a completely classified sample.
\item \texttt{Parameters:} Estimated parameters:which include $\bm{\pi}$, $\bm{\mu}$, $\bm{\Sigma}$ and $\bm{\xi}$ (relevant only if \texttt{ModelType = 'full'}).
\end{itemize}
 
\section{Example: Gastrointestinal lesions data}
\label{sec:example}

In this section, we demonstrate the practical application of the \textsf{gmmsslm} package using the gastrointestinal lesions data from \citet{mesejo2016computer}. The dataset comprises 76 colonoscopy videos, the histology (classification ground truth), and the opinions of the endoscopists (four experts and three beginners). White light and narrow band imaging methods were used to classify whether the lesions were benign or malignant. Each of the $n=76$ observations consists of four selected features extracted from the colonoscopy videos. A panel of seven endoscopists viewed the videos to give their opinion as to whether each patient needed resection (malignant) or no-resection (benign). We formed our partially classified sample as follows. Feature vectors for which all seven endoscopists agreed were taken to be classified with labels specified either as 1 (resection) or 2 (no-resection) using the ground-truth labels. Observations for which there was not total agreement among the endoscopists were taken as having missing labels, denoted by \texttt{NA}.

\begin{figure}[ht]
\centering
\includegraphics[width=10cm]{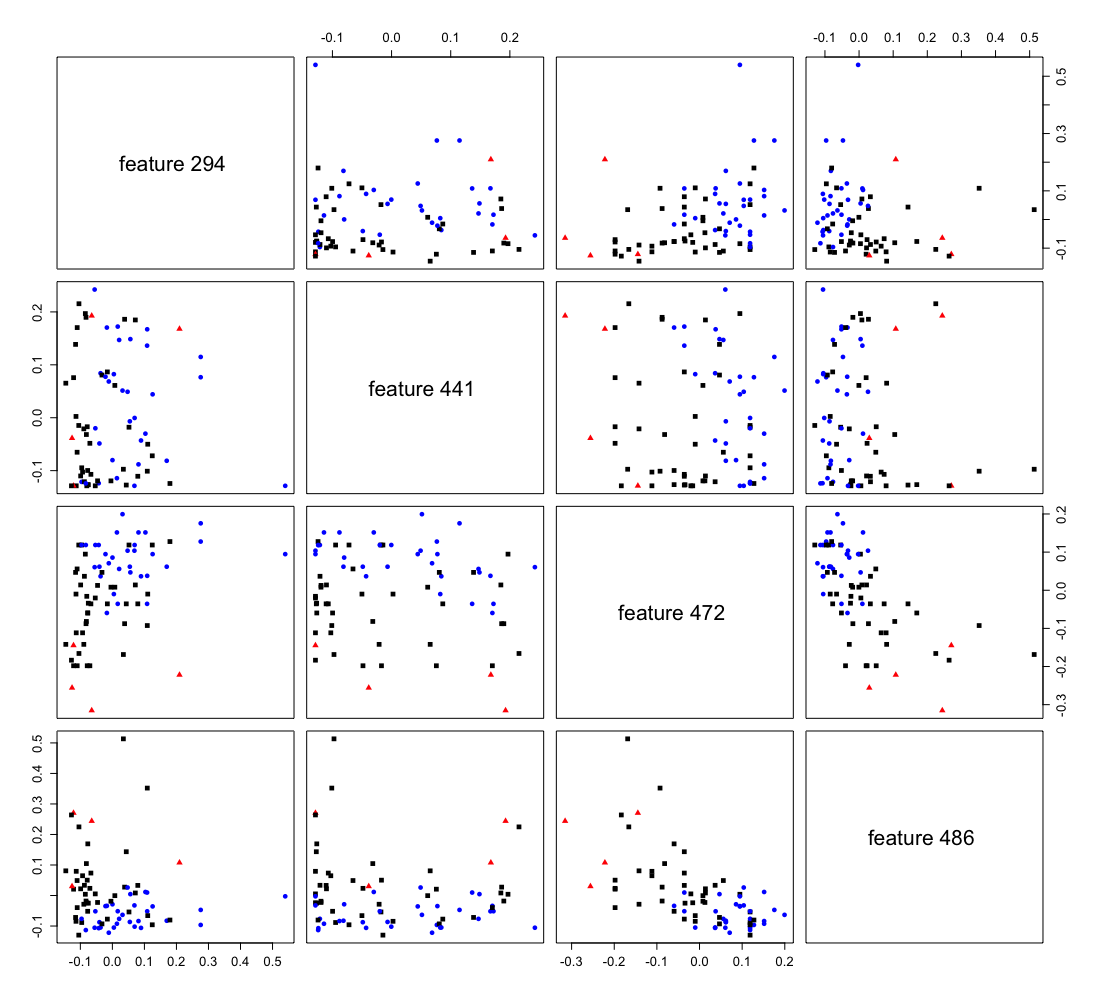}
\caption{Gastrointestinal dataset. The black squares correspond to the unlabelled observations, the red triangles denote the benign observations and the blue circles denote the malignant observations. Observations are treated as unlabelled if fewer than seven endoscopists assigned the same class label to the feature vector.}
\label{fig:dataset}
\end{figure}

We begin by loading the \textsf{gmmsslm} package. The \texttt{gastro\_data} dataset from \textsf{gmmsslm} is then loaded. The feature matrix is denoted as \texttt{X}, where \texttt{n} represents the total number of observations and \texttt{p} their dimensionality. We also define \texttt{g = 2} as the number of classes.
\begin{verbatim}
library(gmmsslm)
data('gastro_data')
X <- as.matrix(gastro_data[,1:4])
n <- nrow(X)
p <- ncol(X)
g <- 2
\end{verbatim}
An initial pair plot is constructed to visualise the locations of the labelled and unlabelled data.
\begin{verbatim}
classagree <- ifelse(is.na(gastro_data$class_agreement), 3,
                gastro_data$class_agreement)
shapevec <- plyr::mapvalues(unclass(classagree), 1:3, 
          to = c(16, 17, 15)) 
colvec <- plyr::mapvalues(unclass(classagree), 1:3,
	          to = c('BLUE', 'RED', 'BLACK'))
pairs(X, pch = shapevec, col = colvec)
\end{verbatim}
Figure~\ref{fig:dataset} shows a plot of the data with the class labels of the feature vectors.
The black squares denote the unlabelled observations, red triangles denote the benign observations and blue circles denote the malignant observations. The unlabelled observations tend to be located in regions of class overlap.

\begin{figure}[ht]
\centering
\includegraphics[width=10cm]{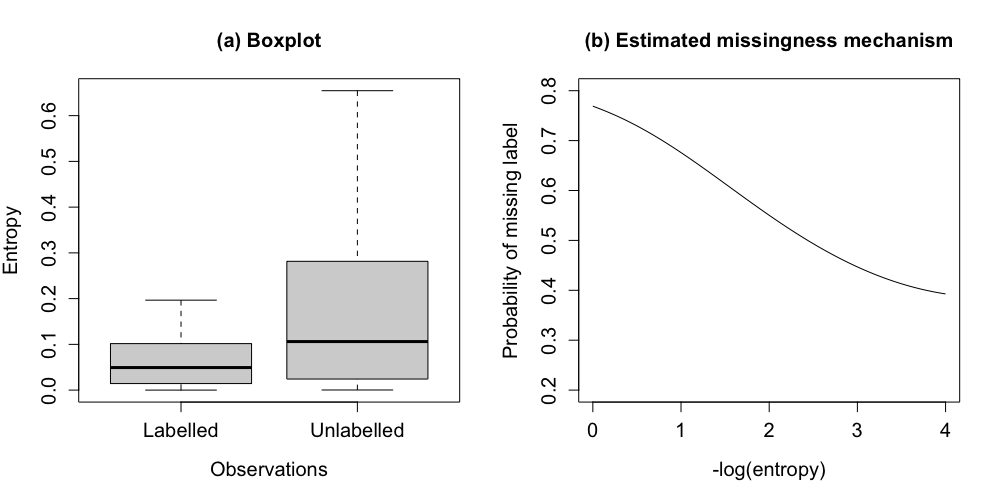}
\caption{Analysing the gastrointestinal dataset concerning the relationship between entropy and observations, both labelled and unlabelled.}
\label{fig:entropy}
\end{figure}

We can now verify if the incomplete data is appropriate for modelling with our proposed approach. Specifically, we need to ascertain whether there exists a missing-data mechanism in the missing labels. The function \texttt{plot\_missingness()} is employed for this purpose, producing a box plot of the initial estimated entropies for labelled and unlabelled groups. Furthermore, it offers a plot showcasing a Nadaraya-Watson kernel estimate for the probability of missing labels against the negative log entropy. The \texttt{plot\_missingness()} function requires initial estimates of $\bm{\pi}$, $\bm{\mu}$ and $\bm{\Sigma}$. To acquire these estimates, we utilise \texttt{initialvalue(dat, zm, g, ncov)}. Note that the \texttt{ncov} parameter specifies the structure of \texttt{sigma}. By default, \texttt{ncov = 2}. When set to \texttt{ncov = 1}, it denotes a common covariance matrix, whereas \texttt{ncov = 2} signifies unequal covariance matrices.
\begin{verbatim}
zm <- gastro_data$class_agreement
inits <- initialvalue(dat = X, zm = zm, g = g, ncov = 2)
plot_missingness(dat = X, g = 2, parlist = inits, zm = zm, 
        bandwidth = 3, range.x = c(0, 4), ylim = c(0.2, 0.8), 
        kernel = 'normal')
\end{verbatim}
Figure~\ref{fig:entropy}(a) compares the box plots of the estimated entropies in the labelled and unlabelled groups. Figure~\ref{fig:entropy}(b) presents the Nadaraya-Watson kernel estimate of the probability of missing labels.
From Figure~\ref{fig:entropy}(a), we find that the unlabelled observations typically have higher entropy than the labelled observations. Figure~\ref{fig:entropy}(b) shows that the estimated probability of a missing class label increases as the log entropy increases. This relation is in accordance with \eqref{eq:qfun}. The higher the entropy of a feature vector, the higher the probability of its class label being unavailable.
 
To process the data with the Gaussian mixture model with a missing-data mechanism, we utilise the \texttt{gmmsslm()} function and set \texttt{type = 'full'}. Prior to this, initial values for $\xi_0$ and $\xi_1$ are essential. We leverage the \texttt{glm()} function from the base package \textsf{stats} as follows:
\begin{verbatim}
en <- get_entropy(dat = X, n = n, p = p, g = g, 
    paralist = inits)
m <- ifelse(gastro_data$missinglabel_indicator == 1, 0, 1)
xi_inits <- coef(glm(m ~ en, family = 'binomial'))
fullfit <- gmmsslm(dat = X, zm = zm, paralist = inits, 
    xi = xi_inits, type = 'full')
\end{verbatim}
The \texttt{gmmsslm()} function yields a \texttt{gmmsslm} object as its output. To extract a comprehensive summary from the fitted model, we can employ the \texttt{summary()} function. This summary reports various aspects, including the likelihood value, fitted variance structure, convergence status, number of iterations, total number of observations, dimension, model type and estimated parameters.
\begin{verbatim}
summary(fullfit)

Table:
                  [,1]                         
Likelihood        "-244.9371"                  
VarianceStructure "Unequal covariance matrices"
Convergence       "1"                          
Iteration         "205"                        
TotalObservation  "76"                         
Dimension         "4"                          
ModelType         "full"                       

Parameters:

$pi
[1] 0.7758885 0.2241115

$mu
             [,1]        [,2]
[1,]  0.018823669 -0.06329461
[2,]  0.002130841 -0.00751177            
[3,]  0.050076855 -0.16765507         
[4,] -0.042663822  0.14306282

$sigma
, , 1

              [,1]          [,2]         [,3]          [,4]
[1,]  1.302248e-02  0.0002572603  0.002515218 -8.461135e-05
[2,]  2.572603e-04  0.0128121495 -0.001279286 -6.947834e-04
[3,]  2.515218e-03 -0.0012792861  0.005025436 -1.812703e-03
[4,] -8.461135e-05 -0.0006947834 -0.001812703  3.683971e-03

, , 2

              [,1]          [,2]          [,3]         [,4]
[1,]  0.0086012175  0.0009785018 -0.0004327248  0.004907910
[2,]  0.0009785018  0.0147164175 -0.0051017162 -0.003288298
[3,] -0.0004327248 -0.0051017162  0.0039913872  0.001410450
[4,]  0.0049079101 -0.0032882976  0.0014104502  0.018480950

$xi
(Intercept)          en 
  1.8276442   0.1434896                             
\end{verbatim}
The initial values have minimal impact on the ECM algorithm's outcomes in this instance. For further exploration of this aspect, interested readers can refer to the online supplementary materials containing relevant code for their own sensitivity analysis. Additionally, we can utilise the \texttt{predict()} function to obtain predicted labels derived from the fit by \texttt{gmmsslm()}. This function returns the predicted labels for the unclassified data initially input into \texttt{gmmsslm()}.
\begin{verbatim}
> predict(fullfit) 
 [1] 1 1 1 1 1 1 1 1 1 2 1 2 1 1 2 1 1 2 1 1 1 1 2 2 2 2
 [27] 2 2 1 2 1 2 1 2 2 1 1 1 1 1 1
\end{verbatim}
Subsequently, we use the function \texttt{paraextract()} to extract the parameter list from the fit by \texttt{gmmsslm()}, then we compute the conditional error rate for the unlabelled data.
\begin{verbatim}
X_ul <- X[m == 1,]
labeltruth = vector(length = n)
labeltruth[gastro_data$`ground truth` == 'resection'] <- 1
labeltruth[gastro_data$`ground truth` == 'no-resection'] <- 2
clust_ul <- labeltruth[m == 1]
err_ful <- erate(dat = X[m == 1,], p, g,
    paralist = paraextract(fullfit), clust = clust_ul)
err_ful
[1] 0.2215959
\end{verbatim}

We now use \texttt{gmmsslm()} to compare the performance of the Bayes' classifier with the unknown parameter vector $\bm{\theta}$ estimated by $\hat{\bm{\theta}}_\mathrm{CC}$, $\hat{\bm{\theta}}_\mathrm{PC}^{(\mathrm{ig})}$ and $\hat{\bm{\theta}}_\mathrm{PC}^{(\mathrm{full})}$. Error rates are estimated using leave-one-out cross-validation.
\begin{verbatim}
clust_com <- clust_ign <- clust_ful <- numeric(n)
for(i in 1:n){
  comfit <- gmmsslm(dat = X[-i,], zm = labeltruth[-i],
            paralist = inits, type = 'com')
  ignfit <- gmmsslm(dat = X[-i,], zm = zm[-i],
            paralist = inits, type = 'ign')
  fullfit <- gmmsslm(dat = X[-i,], zm = zm[-i],
            paralist = inits, xi = xi_inits, type = 'full')
  clust_com[i] <- bayesclassifier(dat = X[i,], p, g,
                paralist = paraextract(comfit))
  clust_ign[i] <- bayesclassifier(dat = X[i,], p, g,
                paralist = paraextract(ignfit))
  clust_ful[i] <- bayesclassifier(dat = X[i,], p, g,
                 paralist = paraextract(fullfit))
}
1 - mean(clust_com == labeltruth)
1 - mean(clust_ign == labeltruth)
1 - mean(clust_ful == labeltruth)
\end{verbatim}
The estimated conditional error rates are reported in Table~\ref{tab:errorrate}.
\begin{table}[ht]
\centering
\caption{Results for the Gastrointestinal dataset. The dataset has $n=76$ observations, $p=4$ features and $g=2$ classes.}
\begin{tabular}{cccc}
\toprule
& $n_\mathrm{c}$ (classified) & $n_\mathrm{uc}$ (unclassified) & Error rate \\
\midrule
$R(\hat{\bm{\theta}}_\mathrm{PC}^{(\mathrm{full})})$ & 35 & 41 & 0.158 \\
$R(\hat{\bm{\theta}}_\mathrm{PC}^{(\mathrm{ig})})$ & 35 & 41 & 0.211 \\
$R(\hat{\bm{\theta}}_\mathrm{CC})$ & 76 & 0 & 0.171 \\
\bottomrule
\end{tabular}
\label{tab:errorrate}
\end{table}
The classifier based on the estimates of the parameters using the full likelihood for the partially classified training sample has lower estimated error rate than that of the rule that would be formed if the sample were completely classified.

\section{Summary}
\label{sec:summary}

The \textsf{R} package \textsf{gmmsslm} implements the semi-supervised learning approach proposed by \cite{ahfock2020apparent} for estimating the Bayes' classifier from a partially classified training sample in which some of the feature vectors have missing labels. It uses a generative model approach whereby the joint distribution of the feature vector and its ground-truth label is adopted. Each of $g$ pre-specified classes to which a feature vector can belong has the multivariate Gaussian distribution. The conditional probability that a feature vector has a missing label is formulated in a framework in which the missingness mechanism models this probability to depend on the entropy of the feature vector using a logistic model. The parameters in the Bayes' classifier are estimated by ML via an ECM algorithm. The package applies to classes with equal or unequal covariance matrices in their multivariate Gaussian distributions. In application to a real-world medical dataset, the estimated error rate of the Bayes' classifier based on the partially classified training sample is lower than that of the Bayes' classifier formed from a completely classified sample.

\begin{Appendix}
\label{app:theory}

Under the model \eqref{eq:qfun} for MAR labels in the case of the two-class homoscedastic Gaussian model, \citet{ahfock2020apparent} derived the following theorem that motivates the development of a package to implement this semi-supervised learning approach for possibly multiple classes with multivariate Gaussian distributions.
\begin{theorem}
The Fisher information about $\bm{\beta}$ in the partially classified sample $\bm{x}_\mathrm{PC}$ via the full likelihood function $L_\mathrm{PC}^{(\mathrm{full})}(\bm{\Psi})$ can be decomposed as
\begin{equation}
\label{eq:thrm}
\bm{I}_\mathrm{PC}^{(\mathrm{full})}(\bm{\beta})=\bm{I}_\mathrm{CC}(\bm{\beta})-\gamma(\bm{\Psi})\bm{I}_\mathrm{CC}^{(\mathrm{clr})}(\bm{\beta})+\bm{I}_\mathrm{PC}^{(\mathrm{miss})}(\bm{\beta}),
\end{equation}
where $\gamma(\bm{\Psi})$ is the proportion of missing labels, $\bm{I}_\mathrm{CC}(\bm{\beta})$ is the information about $\bm{\beta}$ in the completely classified sample $\bm{x}_\mathrm{CC}, \bm{I}_\mathrm{CC}^{(\mathrm{clr})}(\bm{\beta})$ is the conditional information about $\bm{\beta}$ under the logistic regression model fitted to the class labels in $\bm{x}_\mathrm{CC}$ and $\bm{I}_\mathrm{PC}^{(\mathrm{miss})}(\bm{\beta})$ is the information about $\bm{\beta}$ in the missing-label indicators under the assumed logistic model for their distribution given their associated features in the partially classified sample $\bm{x}_\mathrm{PC}$.
\end{theorem}

The expression \eqref{eq:thrm} for the Fisher information about the vector of discriminant function coefficients contains the additional term $\bm{I}_{\mathrm{PC}}^{(\mathrm{miss})}(\bm{\beta})$, arising from the additional information about $\bm{\beta}$ in the missing-label indicators $m_j$. This term has the potential to compensate for the loss of information in not knowing the true labels of those unclassified features in the partially unclassified sample. The compensation depends on the extent to which the probability of a missing label for a feature depends on its entropy. It follows that if
\begin{equation}
\label{eq:inequality}
\bm{I}_{\mathrm{PC}}^{(\mathrm{miss})}(\bm{\beta})>\gamma(\bm{\Psi})\bm{I}_{\mathrm{CC}}^{(\mathrm{clr})}(\bm{\beta}),
\end{equation}
there is an increase in the information about $\bm{\beta}$ from the partially classified sample over the information $\bm{I}_{\mathrm{CC}}(\bm{\beta})$ from the completely classified sample. The inequality in \eqref{eq:inequality} is used to mean that the difference of the left- and right-hand sides of the inequality is a positive definite matrix.

The reader is referred to (36) in \cite{ahfock2020apparent} for a precise definition of the conditional information term $\bm{I}_{\mathrm{CC}}^{(\mathrm{clr})}(\bm{\beta})$. By deriving the asymptotic relative efficiency of the Bayes' rule using the full ML estimate of $\bm{\beta}$, \citet{ahfock2020apparent} showed that the asymptotic expected excess error rate using the partially classified sample $\bm{x}_\mathrm{PC}$ can be much lower than the corresponding excess rate using the completely classified sample $\bm{x}_\mathrm{CC}$. The contribution to the Fisher information from the missingness mechanism can be relatively high if $\xi_{1}$ is large as the location of the unclassified features in the feature space provides information about regions of high uncertainty and, hence, where the entropy is high.

However, when extending to multiple classes with multivariate Gaussians, providing a similar asymptotic analysis becomes considerably more complex. We are inclined to believe that the foundational ideas of Theorem 1 should hold for the general case. However, a comprehensive analytical verification would be extensive and is beyond the scope of the current manuscript. We recognise the importance of such a verification and are planning to address this in our future research, possibly by leveraging numerical analysis techniques.

\end{Appendix}

\begin{Appendix}
\label{app:discriminant}

Since the expression for the optimal error rate is complicated, it is difficult to give a theoretical analysis of likelihood inference based on a missing-data mechanism in the general case. In the particular case of $g=2$ classes and under the assumption of equal covariance $\bm{\Sigma}_1=\bm{\Sigma}_2=\bm{\Sigma}$, \cite{ahfock2020apparent} provided a Taylor series approximation for the logarithm of entropy as
\begin{equation*}
\log\left\{e\left(\bm{y}_j;\bm{\theta}\right)\right\}=-\left[\log \{\log (2)\}+d^2\left(\bm{y}_j;\bm{\beta}\right)/\{8\log(2)\}\right]+O\left(d^4\left(\bm{y}_j;\bm{\beta}\right)\right),
\end{equation*}
where
\begin{equation*}
d\left(\bm{y}_j;\bm{\beta}\right)=\beta_0+\bm{\beta}_1^\top \bm{y}_j,
\end{equation*}
with
\begin{equation*}
\beta_0=\log\left(\pi_1/\pi_2\right)-\frac{1}{2}\left(\bm{\mu}_1+\bm{\mu}_2\right)^\top \bm{\Sigma}^{-1}\left(\bm{\mu}_1-\bm{\mu}_2\right)\quad\text{ and }\quad\bm{\beta}_1=\bm{\Sigma}^{-1}\left(\bm{\mu}_1-\bm{\mu}_2\right).
\end{equation*}

This expression provides a linear relationship between the negative log entropy $\log\{e(\bm{y}_j;\bm{\theta})\}$ and the square of the discriminant function $d^2(\bm{y}_j;\bm{\beta})$. Therefore, the negative log entropy in the conditional probability \eqref{eq:qfun}, can be replaced by the square of the discriminant function,
\begin{equation*}
q(\bm{y}_j;\bm{\beta},\bm{\xi})=\frac{\exp\{\xi_0-\xi_1d^2(\bm{y}_j;\bm{\beta})\}}{1+\exp\{\xi_0-\xi_1 d^2(\bm{y}_j;\bm{\beta})\}}.
\end{equation*}
The missing-label indicator based on the entropy ($g>2$) or the square of the discriminant function ($g=2$) can be generated using the function \texttt{rlabel(dat, pi, mu, sigma, xi)}. The element of the outputs represent a missing label when equal to 1 and an available label when equal to 0.

\end{Appendix}

\begin{Appendix}
\label{app:ecm}

We apply the ECM algorithm of \citet{meng1993maximum} to compute the ML estimate of $\bm{\Psi}$ on the basis of the full likelihood $L_\mathrm{PC}^{(\mathrm{full})}(\bm{\Psi})$. The adopted ECM framework makes the obvious choice of declaring the `missing' data to be the missing labels $\bm{z}_j$ for those features $\bm{y}_j$ with $m_j=1$.

\textbf{E step:} It handles the presence of the introduced missing data by forming on the $(k+1)$th iteration the so-called $Q$-function $Q(\bm{\Psi}; \bm{\Psi}^{(k)})$ equal to the expectation of the complete data log likelihood conditional on the observed data $\bm{y}$, using the current estimate $\bm{\Psi}^{(k)}$ for $\bm{\Psi}$. As this complete data log likelihood is linear in the missing class labels, this expectation conditional on $\bm{y}$ is effected by replacing the unobservable $z_{ij}$ by its conditional expectation given $\bm{y}, z_{ij}^{(k)}$, where
\begin{equation*}
\begin{split}
z_{ij}^{(k)}&=\E_{\bm{\Psi}^{(k)}}\{Z_{ij}\mid\bm{y}\} \\
&={\Pr}_{\bm{\Psi}^{(k)}}\{Z_{ij}=1\mid\bm{y}\} \\
&=\tau_i(\bm{y}_j;\bm{\Psi}^{(k)}) \\
&=\frac{\pi_i^{(k)}\phi(\bm{y}_j;\bm{\mu}_i^{(k)},\bm{\Sigma}_i^{(k)})}{\sum_{h=1}^g\pi_h^{(k)}\phi(\bm{y}_j;\bm{\mu}_h^{(k)},\bm{\Sigma}_h^{(k)})}.
\end{split}
\end{equation*}
Accordingly, we have that
\begin{equation*}
\begin{split}
Q(\bm{\Psi};\bm{\Psi}^{(k)})&=\sum_{j=1}^n(1-m_j)\sum_{i=1}^g z_{ij} \{\log \pi_i+\phi(\bm{y}_j;\bm{\mu}_i,\bm{\Sigma}_i)\} \\
&\quad+\sum_{j=1}^nm_j\sum_{i=1}^g z_{ij}^{(k)} \{\log \pi_i+\phi(\bm{y}_j;\bm{\mu}_i,\bm{\Sigma}_i)\} \\
&\quad+\sum_{j=1}^n\left[(1-m_j)\log\{1-q(\bm{y}_j;\bm{\theta},\bm{\xi})\}+m_j\log q(\bm{y}_j; \bm{\theta},\bm{\xi})\right].
\end{split}
\end{equation*}

We calculate the updated value $\bm{\Psi}^{(k+1)}$ of $\bm{\Psi}$ using two conditional maximisation (CM) steps.

\textbf{CM-step1:} We fix $\bm{\xi}$ at its current value $\bm{\xi}^{(k)}$ and update $\bm{\theta}$ to $\bm{\theta}^{(k+1)}$ given by
\begin{equation*}
\bm{\theta}^{(k+1)}=\underset{\bm{\theta}}{\arg\max}\,Q(\bm{\theta},\bm{\xi}^{(k)};\bm{\theta}^{(k)},\bm{\xi}^{(k)}),
\end{equation*}
where
\begin{equation*}
\begin{split}
Q(\bm{\theta},\bm{\xi}^{(k)};\bm{\theta}^{(k)},\bm{\xi}^{(k)})&=\sum_{j=1}^n(1-m_j)\sum_{i=1}^gz_{ij}^{(k)}\{\log\pi_i+\log f_i(\bm{y}_j;\bm{\omega}_i)\} \\
&\quad+\sum_{j=1}^nm_j\sum_{i=1}^gz_{ij}^{(k)}\{\log\pi_i+\log f_i(\bm{y}_j;\bm{\omega}_i)\} \\
&\quad+\sum_{j=1}^n\left[(1-m_j)\log\{1-q(\bm{y}_j;\bm{\theta},\bm{\xi}^{(k)})\}+m_j\log q(\bm{y}_j;\bm{\theta},\bm{\xi}^{(k)})\right].
\end{split}
\end{equation*}

\textbf{CM-step 2:} We now fix $\bm{\theta}$ at its updated value $\bm{\theta}^{(k+1)}$ and update $\bm{\xi}$ to $\bm{\xi}^{(k+1)}$ as
\begin{equation*}
\bm{\xi}^{(k+1)}=\underset{\bm{\xi}}{\arg\max}\,Q(\bm{\theta}^{(k+1)},\bm{\xi};\bm{\theta}^{(k)},\bm{\xi}^{(k)}),
\end{equation*}
which reduces to
\begin{equation*}
\bm{\xi}^{(k+1)}=\underset{\bm{\xi}}{\arg\max}\,\log L_{\mathrm{PC}}^{(\mathrm{miss})}(\bm{\theta}^{(k+1)},\bm{\xi}),
\end{equation*}
on retaining only terms that depend on $\bm{\xi}$.

As $L_\mathrm{PC}^{(\mathrm{miss})}(\bm{\theta}^{(k+1)},\bm{\xi})$ belongs to the regular exponential family, we use the function \texttt{glm()}. The estimate $\bm{\Psi}_{\mathrm{PC}}^{(\mathrm{full})}$ is given by the limiting value of $\bm{\Psi}^{(k)}$ as $k$ tends to infinity. We take the ECM algorithm as having converged when
\begin{equation*}
\log L_{\mathrm{PC}}^{(\mathrm{full})}(\bm{\Psi}^{(k+1)})-\log L_{\mathrm{PC}}^{(\mathrm{full})}(\bm{\Psi}^{(k)})
\end{equation*}
is less than some arbitrarily specified value.

\end{Appendix}

\bibliographystyle{anzsj}
\bibliography{references}

\end{document}